\documentclass[bibyear]{aa}
\usepackage{graphicx}
\usepackage[varg]{txfonts}
\usepackage{hyperref}
\pdfoutput=1
\begin{document}

\title{What drives galactic magnetism?\thanks{Based on
observations with the 100-m telescope at Effelsberg
operated by the Max-Planck-Institut f\"ur Radioastronomie (MPIfR) on behalf
of the Max-Planck-Gesellschaft.
}
}
\author{K.\,T.~Chy\.zy\inst{1}
\and S.\,S.~Sridhar\inst{2,3}
\and W.~Jurusik\inst{1}
}
\offprints{Krzysztof T. Chy\.zy, \email{krzysztof.chyzy@uj.edu.pl}}
\institute{Astronomical Observatory of the Jagiellonian University, ul. Orla 171, 30-244 Krak\'ow, Poland\label{inst1}
\and Kapteyn Astronomical Institute, University of Groningen, Postbus 800, 9700AV Groningen, The Netherlands.
\and ASTRON, the Netherlands Institute for Radio Astronomy, Postbus 2, 7990 AA, Dwingeloo, The Netherlands\label{inst2}
}

\date{Received date/ Accepted date}

\abstract
{}
{Magnetic fields are important ingredients of the interstellar medium.
They are suspected to be maintained by dynamo processes related to
star-formation activity, properties of the interstellar medium and
global features of galaxies. We aim to use statistical analysis of a large number of various
galaxies to probe, model, and understand
relations between different galaxy properties and magnetic fields.
}
{
We have compiled a sample of 55 galaxies including low-mass dwarf and
Magellanic-types, normal spirals and several massive starbursts, and applied
principal component analysis (PCA) and regression methods to
assess the impact of various galaxy properties on the observed magnetic fields.
}
{
According to PCA the global galaxy parameters (like \ion{H}{I}, H$_2$, and dynamical mass,
star formation rate (SFR), near-infrared luminosity, size, and rotational velocity) are all mutually
correlated and can be reduced to a single principal component. Further PCA performed for
global and intensive (not size related) properties of galaxies (such as gas density, and surface density
of the star formation rate, SSFR), indicates that magnetic field strength $B$ is connected mainly
to the intensive parameters, while the global parameters have only weak relationships with $B$.
We find that the tightest relationship of $B$ is with SSFR, which is described by a power-law with
an index of $0.33\pm 0.03$. The relation is observed for galaxies with the global SFR spread
over more than four orders of magnitude. Only the radio faintest dwarf galaxies deviate from
this relation probably due to the inverse Compton losses of relativistic electrons or
long turbulence injection timescales.
The observed weaker associations of $B$ with galaxy dynamical mass and the
rotational velocity we interpret as indirect ones, resulting from the observed connection of
the global SFR with the available total H$_2$ mass in galaxies.
Using our sample we constructed a diagram of $B$ across the Hubble sequence which reveals
that high values of $B$ are not restricted by the Hubble type and even dwarf (starbursting)
galaxies can produce strong magnetic fields. However, weaker fields appear exclusively in
later  Hubble types and $B$ as low as about $5\,\mu$G is not seen among typical spirals.
}
{
The processes of generation of magnetic field in the dwarf and Magellanic-type galaxies are similar
to those in the massive spirals and starbursts and are mainly coupled to local star-formation activity
involving the small-scale dynamo mechanism.
}


\maketitle

\section{Introduction}
\label{s:intro}

The interstellar medium (ISM) is pervaded with magnetic fields of energy similar
to other ISM species, generated by small-scale and large-scale ($\alpha-\Omega$)
dynamo processes and transported with the bulk motion of interstellar plasma (Beck \cite{beck16}).
Observational evidence suggests that magnetic fields in galaxies play an important role 
in regulating the ISM by confining cosmic ray electrons (Berezinskii et al.~\cite{berezinskii90}) 
and providing vertical support to the interstellar gas (Fletcher \& Shukurov~\cite{fletcher01}), 
and regulating angular momentum transfer in gas clouds that eventually collapse to form stars 
(Zweibel \& Heiles~\cite{zweibel97}). A study of individual nearby galaxies provides us with data
on topology and strength of magnetic fields in various galactic environments.
Revealing statistical relations between the various observational parameters to describe
the galaxies and the magnetic field strength is a helpful tool in recognising, modelling,
and understanding the impact of various physical processes involved in the formation and evolution 
of magnetic fields in the galaxies.

There are objective obstacles encountered in such studies, like difficulties in observing
optically and radio-weak dwarf galaxies and distant protogalaxies. For example,
a systematic study of low-mass galaxies in the Local Group revealed surprisingly little
information concerning magnetic fields in these objects as only three out of 12 dwarfs were detected
in the radio domain (Chy\.zy et al.~\cite{chyzy11}).
The results obtained indicated that magnetic fields in the dwarf galaxies
are rather weak, with a mean value of total field strength of only $4\,\mu$G.
Basing on the radio-detected low-mass galaxies (from the Local Group as well as from outside it)
a power-law relation of the magnetic field strength and the surface density of star formation
rate (SSFR) with an index of $0.30\pm 0.04$ was determined. Some other relationships
of magnetic fields with galaxy parameters were also found.
To what extent the relationships obtained for the low-mass galaxies remain valid also
for the massive galaxies and starbursts, is not known.

Recently, Tabatabaei et al. (\cite{taba16})
indicated that the large-scale (ordered) magnetic field  in a sample of 26 galaxies is
proportional to their rotational speed. The enhanced field in this case could be due
to gas compression and shearing flows in fast rotating systems. In another work, 
Van Eck et al.~(\cite{vaneck15}) used 20 well-observed nearby galaxies to present a statistically
important relation of the total magnetic field strength with the SSFR (with the power-law
index $n=0.19\pm 0.03$) as well as with the density of molecular gas ($n=0.21\pm 0.04$).
The magnetic pitch angle appeared to be associated with the total gas density, star formation rate,
and strength of the axisymmetric component of the large-scale part of magnetic field.
A steeper relation between the total field and the SSFR was found by Heesen et
al. (\cite{heesen14}) for 17 galaxies, containing two dwarfs. Performing similar studies for
a much larger sample of different galaxies is much needed.

In order to investigate importance of various correlations of observed parameters of galaxies,
Disney et al. (\cite{disney08}) used the principal component analysis (PCA) to statistically
analyse a sample of 200 galaxies, showing that the galaxies  can be described
in a much simpler way than suggested by the hierarchical structure formation theory
and are actually controlled by a small number of dominating parameters.
In later studies, Li \& Mao (\cite{li13}) reproduced the results of Disney et al. for a sample 
of 2000 SDSS galaxies and used PCA to construct parameters to better differentiate the galaxies
than the original observables, like colour, stellar age, or stellar mass. They also
proved that the galaxy environment did not affect galaxy morphology to a greater extent, while
significantly changing galactic colours.

In this paper, we explore how the statistical relationships determined for the low-mass objects concern 
the general population of galaxies, probing relations of magnetic field with a number of properties 
describing galaxies in a sample of 55 objects. Our sample includes faint dwarf galaxies, normal spirals, 
and several massive starbursts, in order to cover a wide range of star formation processes
and to find out possible interrelations for all the objects. We use our radio observations of low-mass objects and acquire
information on the other galaxies from the available publications. The sample's size allows us to inspect magnetic fields
across the Hubble sequence. The radio-faintest dwarf galaxies, for which stacking experiments of their radio maps
were performed, are also analysed. The investigation involves a statistical analysis of the galaxy sample basing 
on two methods, PCA and regression modelling.

\section{Galaxy sample}
\label{s:sample}

\subsection{Low-mass objects}
\label{s:lowmasssample}

In our low-mass sample, we included low-mass galaxies from our radio observations made with the 100-m Effelsberg
telescope: three dwarf galaxies from Chy\.zy et al. (\cite{chyzy11}) observed at 2.64\,GHz (NGC\,6822, IC\,10,
IC\,1613), five low-mass, Magellanic-type galaxies observed at 4.85\,GHz and/or 8.35\,GHz (NGC\,3239, NGC\,4027,
NGC\,4618, NGC\,5204, UGC\,11861), peculiar, `pure disk' objects (NGC\,2976 and NGC\,4605) (Jurusik
et al. \cite{jurusik14}), as well as three galaxies (NGC\,4236, NGC\,4656, IC\,2574) from Chy\.zy et al. \cite{chyzy07}. For
all these galaxies we calculated the  total magnetic field strength $B$ assuming energy equipartition between magnetic
fields and cosmic rays (Beck \& Krause \cite{beck05}). The separation of thermal emission from the radio total
flux was achieved with the help of H$\alpha$ fluxes. In the case of Magellanic and peculiar objects, we corrected
the H$\alpha$ fluxes for dust attenuation using information on the infrared (dust) emission (see Jurusik et
al. \cite{jurusik14}). The sizes and masses of these objects are between the dwarf and typical spiral galaxies.

In order to have the best possible representation of radio-faint star-forming dwarf galaxies, we included into the sample UGC\,5456 and
analysed the `common' sample of dwarfs from the stack experiment from Roychowdhury \& Chengalur (\cite{roy12}), while performing a
similar stack experiments for the dwarf galaxies of the Local Group which went undetected in the work of Chy\.zy et al. (\cite{chyzy11}).
Using NVSS (1.4\,GHz) maps for these nine dwarfs from the Local Group (Aquarius, GR 8, WLM, LGS 3, SagDIG, Sextant A, Sextant B, Leo A, and
Pegasus), we were able to estimate only the upper limit of $B=5\pm 1\,\mu$G.
Presumably, the number of our stacked objects was too small for the signal to be detected. Our Effelsberg observations (Chy\.zy et al.
\cite{chyzy11}) at 2.64\,GHz provided a better estimation of this upper limit with $B < 3.8\pm0.6\,\mu$G.

We also added five galaxies from the available work: LMC, SMC, NGC\,4449, NGC\,1569, NGC\,4214. 
The sources of the data for these objects are given in Table~\ref{t:basic}.

\subsection{Massive galaxies}
\label{s:massivesample}

Our sample contained well-researched normal spiral galaxies for which we were able to find proper data in the literature.
To work with the most uniform dataset possible,
we used radio continuum data from the WSRT survey of SINGS galaxies (Braun et al. \cite{braun07}) to estimate the equipartition 
magnetic field strength for 14 objects from the nonthermal emission, taking the thermal fractions from Heesen 
et al. (\cite{heesen14}) and the galaxy inclination  values from HyperLeda or NED.
For other 14 galaxies we used estimations of $B$ (for the entire galaxies) from the compilation of Van Eck (\cite{vaneck15}).
We also added seven well-known spirals from other studies (Table~\ref{t:basic}). Our sample involved  massive starbursts
(NGC\,253, M\,81) as well as luminous infrared radio galaxies (LIRGs: NGC\,3256 and Arp 220).

\subsection{Construction of extensive and intensive parameters}
\label{s:construction}

For each galaxy in the sample, we searched the literature for information on their global properties:
morphological (Hubble) type $T$, inclination $i$, distance $D$, the optical angular radius, which was transformed to the 
linear one $R$, rotational velocity $V$, global SFR, the total \ion{H}{i} mass $M_\mathrm{HI}$,
the total mass of molecular gas $M_\mathrm{H2}$, the near-infrared luminosity $LK$ in $K_s$ band, which is related to the total
galactic stellar mass. We also calculated `tentative' total masses of galaxies, estimating them from the formula: $M\propto R\,V^2$. 
The parameters: SFR, $LK$, $M_{\mathrm{HI}}$, $M_{\mathrm{H2}}$, $M$, $R$ are all extensive properties of galaxies 
and depend on the object size: splitting a galaxy in half would result in decreasing the values of these parameters 
to half of the original ones.  

The mean magnetic field strength, calculated as an average value over the galaxy, is directly related to the 
volume density of magnetic energy and calculated from the radio emission, taking into account the synchrotron pathlength. 
This is an intensive property, independent of galaxy size.
Therefore, we constructed other parameters describing the intensive properties of galaxies, free from the influence of their sizes 
and masses (see e.g. Lara-L{\'o}pez et al. \cite{lara13}). The analysis which global or intensive parameters are mainly 
related to the magnetic field, and which are less important is one of the purposes of our analysis.
We constructed the following set of intensive parameters: the (mean) surface density of star formation rate $\mathrm{SSFR=SFR}/A$, 
the density of hydrogen gas $SM_\mathrm{HI}=M_\mathrm{HI}/A$, the density of H$_2$ gas $SM_\mathrm{H2}=M_\mathrm{H2}/A$, 
near-infrared surface brightness $SLK=LK/A$, where $A$ is the observed surface area of the galaxy. Moreover, we calculated the star 
formation efficiency with reference to the neutral gas $\mathrm{SFE=SFR}/M_\mathrm{HI}$ and the similar efficiency for the H$_2$ 
gas $\mathrm{SFE_{H2}=SFR}/M_\mathrm{H2}$. The intensive parameters involving the magnetic field strength and the surface 
densities are derived for the entire galaxies using their optical
or radio extents. 

We note that in some literature (e.g. Thompson et al. \cite{thompson06}) the magnetic field strength and gas densities
are calculated for restricted regions of strong star formation, which obviously yields different estimates (e.g. in the extreme cases
of M\,82 and Arp\,220, the values of $B$ obtained by us are by an order of magnitude lower than those in Thompson et al. \cite{thompson06}
calculated for compact starbursts).
The main properties of all 55 galaxies are summarised in Table \ref{t:basic}.

\section{Results}
\label{s:results}

\begin{figure}
\centering
\includegraphics[width=0.49\textwidth]{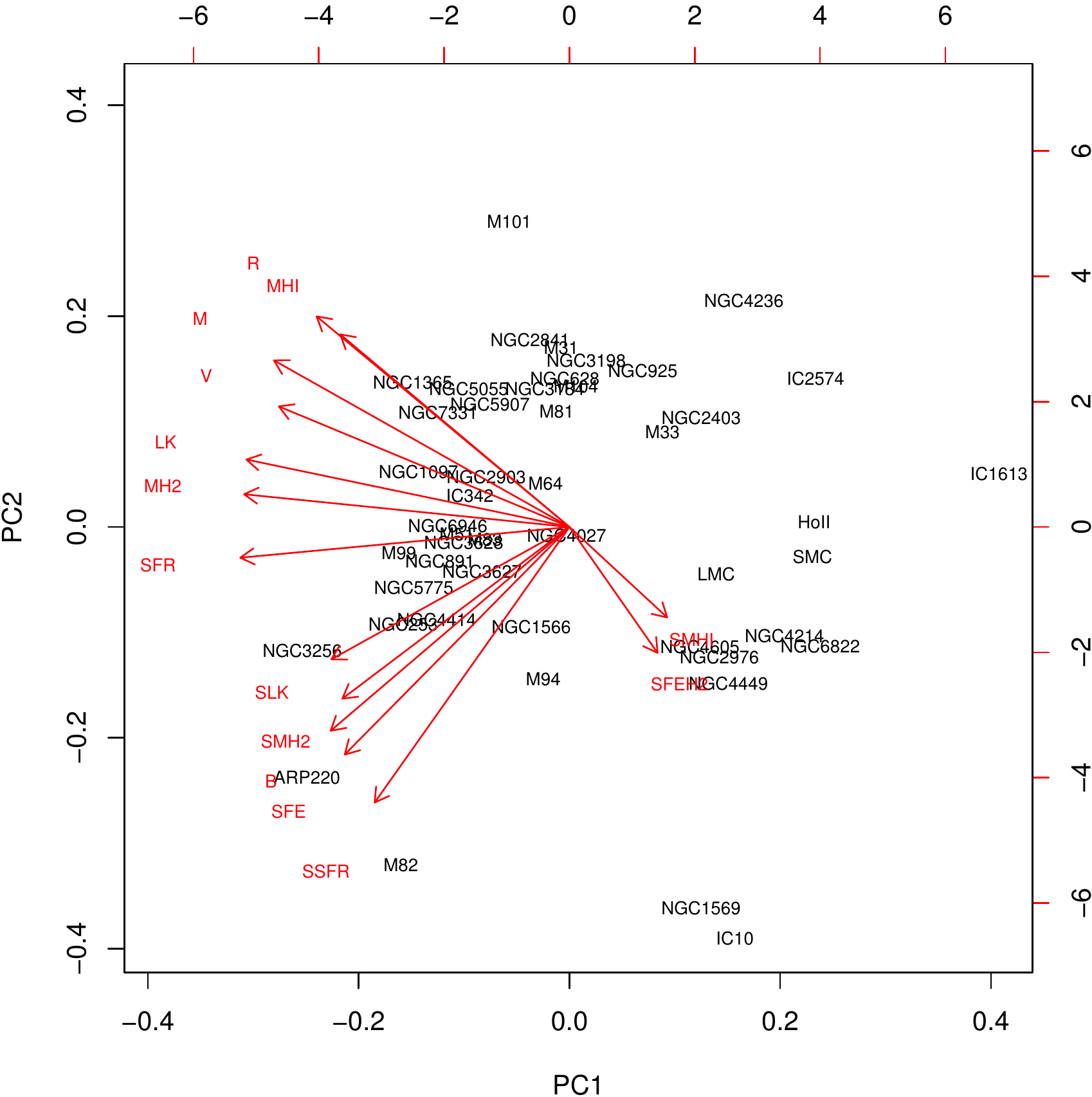}
\caption{Biplot obtained from PCA of all galaxy parameters, showing
the positions of individual galaxies and the directions of the original 
variables (arrows) as
projected into the plane of the first two PCs. The horizontal axis is 
the most varying direction of the
data-set. The positions of galaxies were \emph{scaled down} by the 
standard deviation
of the corresponding PCs multiplied by the square root of the number of 
observations (bottom and
left-hand axes), while the vectors were \emph{scaled up} by the same 
values (top and right-hand axes).
}
\label{f:biplot}
\end{figure}

The investigation of our galaxy sample is performed by applying two statistical methods: 
PCA and two-dimensional regression.

\subsection{Principal component analysis}
\label{s:pca}

\begin{table*}
\caption{Eigenvalues, variances explained by the principal components, and eigenvectors from PCA of global 
parameters and $B$ (see Sect.~\ref{s:pca}). The principal components are denoted as PC1 to PC8. 
Eigenvector components with small ($<$0.1) values indicating little contributions to the principal 
components have been left blank in the table.
}
\begin{center}
\begin{tabular}{lcccccccc}
\hline
\hline
              & PC1 & PC2 & PC3 & PC4 & PC5 & PC6 & PC7 & PC8\\
\hline              
Eigenvalues  &  5.96 & 1.15 & 0.39 & 0.23  & 0.15  & 0.08  & 0.04  & 0.00\\
\hline              
Var. explained &  0.745 & 0.144  & 0.048  & 0.029  & 0.018  & 0.010  & 0.005  & 0.000\\
\hline
$B$                       &    -0.215&  0.767&       &  0.376&       & -0.303& -0.356&         \\
SFR                       &    -0.370&  0.324& -0.263& -0.117& -0.104&       &  0.815&         \\
$M_{\mathrm{HI}}$&    -0.342& -0.305& -0.537&  0.427&  0.480&  0.283&       &         \\
$M_{\mathrm{H2}}$&    -0.368&  0.168&       & -0.768&  0.354&  0.160& -0.309&         \\
$LK$                      &    -0.388&       &  0.187&  0.137& -0.580&  0.647& -0.196&         \\
$R$                       &    -0.360& -0.336& -0.328& -0.129& -0.446& -0.534& -0.203& -0.329  \\
$V$                       &    -0.365& -0.134&  0.644&  0.183&  0.308& -0.110&  0.163& -0.517  \\
$M$                       &    -0.389& -0.228&  0.285&       &       & -0.294&       &  0.790  \\
\hline
\end{tabular}
\end{center}
\label{t:pca1}
\end{table*}

\begin{table*}
\caption{Eigenvalues, variances explained by the principal components, and eigenvectors of PCA of intensive 
parameters and $B$ (see Sect.~\ref{s:pca}). The principal components are denoted as PC1 to PC7. 
Eigenvector components with small ($<$0.1) values indicating little contributions to the principal 
components have been left blank in the table.
}
\begin{center}
\begin{tabular}{lccccccc}
\hline
\hline
              & PC1 & PC2 & PC3 & PC4 & PC5 & PC6 & PC7 \\
\hline              
Eigenvalues  &  3.83 & 1.52 & 0.96 & 0.47  & 0.20  & 0.02 & 0.00\\
\hline              
Var. explained &0.547 & 0.217 & 0.136 & 0.068 & 0.029 & 0.004 & 0.000\\
\hline
 $B$                 & -0.457 &-0.132 & 0.123 & 0.153 & 0.854 &       &         \\
 SSFR                & -0.480 &-0.192 &       & 0.251 &-0.253 &-0.775 &         \\
 SFE                 & -0.471 &       & 0.320 & 0.175 &-0.354 & 0.414 & 0.585        \\
 SFE$_{\mathrm{H2}}$ &        &-0.723 & 0.455 &       &-0.201&  0.208& -0.432       \\
 $SM_{\mathrm{HI}}$  &        &-0.568 &-0.724 & 0.102 &       & 0.226 & 0.301   \\
 $SM_{\mathrm{H2}}$  & -0.428 & 0.302 &-0.369 & 0.206 &-0.198 & 0.357 &-0.617   \\
 $SLK$               & -0.393 &       &       &-0.911 &       &       &         \\
\hline
\end{tabular}
\end{center}
\label{t:pca2}
\end{table*}

PCA is an exploratory technique useful for finding patterns or structure in a multivariate dataset. 
This method combines variables (parameters) that redundantly measure the same property,
and reduces the importance of variables that contribute little information to the data.
It is also useful as a more general statistical tool for describing and understanding the 
data structure. PCA models the covariance or correlation matrix of the data 
to find relationships to best account for the data variance.
As a result, it produces a number of new, statistically independent variables, 
called the principal components (PCs), which are linear combination of the original variables. 

The problem of determining new variables to maximize information (data variance) 
is equivalent to finding eigenvectors and eigenvalues of the data covariance (or correlation) matrix. 
The i-th PC is the line in the data parameter space that follows the eigenvector associated with the i-th
largest eigenvalue measuring the variance in the direction of the i-th PC. Therefore the first PC 
is aligned with the direction of maximum variance in the entire dataset, 
the second one shows the highest variability for all directions orthogonal to the
first PC, and so forth. The number of derived PCs equals the number of the original parameters considered in 
the analysis and the original observations can be expressed in the new coordinates (by projecting onto the PCs). 
We performed such PCA basing on the correlation matrix of logarithmised parameters describing
our sample of galaxies.

In our first PCA approach, we analysed only the global parameters of galaxies
(SFR, $M_{\mathrm{HI}}$, $M_{\mathrm{H2}}$, $LK$, $R$, $V$, and $M$).
It turned out that all the parameters are correlated, allowing for descripting the entire sample
by just one principal component (PC1), which can account for 82\% of variance in the galaxy
parameters. All the global parameters contribute to PC1 to roughly the same extent and with the same sign.
The second and next PCs have eigenvalues smaller than 1 and are considered insignificant.

\begin{table*}
\caption{Parameters of statistical fits.}
\begin{center}
\begin{tabular}{lcccc}
\hline
\hline
Relation            & n[M(Y/X)]          &  n(Bisector)     &  $\rho$/P-value$^a$ & N \\
\hline
$B\propto \mathrm{SFR}^n$    & $0.21\pm 0.02$ & $0.28\pm 0.02$ & 0.68/0.00 & 55\\
$B\propto (M_\mathrm{HI})^n$ & $0.08\pm 0.05$ & $0.63\pm 0.17$ & 0.18/0.18 & 55\\
$B\propto (M_\mathrm{H2})^n$ & $0.15\pm 0.03$ & $0.27\pm 0.03$ & 0.54/0.00 & 48\\
$B\propto (M_\mathrm{gas})^n$ & $0.16\pm 0.04$ & $0.43\pm 0.08$ & 0.38/0.01 & 48\\
$B\propto {LK}^n$     & $0.13\pm 0.03$ & $0.26\pm 0.04$ & 0.49/0.00  & 55\\
$B\propto R^n$      & $0.11\pm 0.08$ & $0.83\pm 0.11$ & 0.16/0.24  & 55\\
$B\propto V^n$      & $0.31\pm 0.09$ & $0.82\pm 0.07$ & 0.35/0.01  & 55\\
$B\propto M\propto (V^2R)^n$ & $0.09\pm 0.03$ & $0.37\pm 0.08$ & 0.30/0.02 & 55\\
\hline
$B\propto \mathrm{SSFR}^n$    & $0.33\pm 0.03$ & $0.41\pm 0.03$ & 0.78/0.00  & 55\\
$B\propto (\mathrm{SSFR_{cor}})^n$ & $0.31\pm 0.03$ & $0.39\pm 0.03$ & 0.80/0.00  & 55 \\
$B\propto (SM_\mathrm{HI})^n$ & $-0.01 \pm 0.09 $ & $-0.96\pm 0.09$ & -0.03/0.85 & 55\\
$B\propto (SM_\mathrm{H2})^n$ & $0.23\pm 0.04$ & $0.37\pm 0.04$ & 0.65/0.00 & 48\\
$B\propto (SM_\mathrm{gas})^n$ & $0.41\pm 0.10$ & $0.82\pm 0.09$ & 0.52/0.00 & 48\\
$B\propto {SLK}^n$     & $0.21\pm  0.04$ & $0.39\pm 0.05$ & 0.60/0.00  & 55\\
$B\propto \mathrm{SFE}^n$    & $0.30\pm 0.03$ & $0.37\pm 0.03$ & 0.75/0.00  & 55\\
$B\propto \left({\mathrm{SFE_{H2}}}\right)^n$ & $0.06\pm  0.07$ & $0.77\pm 0.12$ & 0.07/0.61  & 48\\
\hline
$\mathrm{SFR}\propto (M_\mathrm{HI})^n$   & $0.93\pm 0.13$ & $1.37\pm 0.13$ & 0.65/0.00  & 55\\
$\mathrm{SFR}\propto (M_\mathrm{H2})^n$   & $0.77\pm 0.05$ & $0.86\pm 0.07$ & 0.88/0.00  & 48\\
$\mathrm{SFR}\propto M^n$          & $0.69\pm 0.07$ & $0.94\pm 0.09$ & 0.71/0.00  & 55\\
$\mathrm{SFR}\propto {LK}^n$          & $0.72\pm 0.05$ & $0.85\pm 0.07$ & 0.80/0.00  & 55\\
$\mathrm{SSFR}\propto (SM_\mathrm{HI})^n$ & $0.33\pm 0.23$ & $1.19\pm  0.20$ & 0.15/0.29 & 55\\
$\mathrm{SSFR}\propto (SM_\mathrm{HI})^n$ restr.$^b$ & $0.54\pm 0.21$ & $1.30\pm 0.15$ & 0.30/0.03 & 51\\
$\mathrm{SSFR}\propto (SM_\mathrm{H2})^n$ & $0.67\pm 0.08$ & $0.87\pm 0.10$ & 0.78/0.00 & 48\\
$\mathrm{SSFR}\propto (SM_\mathrm{H2})^n$ restr.$^c$ & $0.96\pm 0.20$ & $1.49\pm 0.18$ & 0.63/0.00 & 27\\
$\mathrm{SSFR}\propto (SM_\mathrm{gas})^n$           & $1.39\pm 0.20$ & $1.94\pm 0.20$ & 0.70/0.00  & 48\\
$\mathrm{SSFR}\propto {SLK}^n$              & $0.56\pm 0.08$ & $0.90\pm 0.10$ & 0.61/0.00  & 55\\
\hline
\hline
\end{tabular}
\end{center}
Notes. $^{(a)}$ -- large P-values mean a low confidence level to reject the hypothesis that the data are not correlated; 
$^{(b)}$ -- restricted so as to not include massive starburst/LIRGs; $^{(c)}$ -- restricted to 
$(3 < SM_\mathrm{H2} < 50)$\,$M_{\sun}\,\mathrm{pc}^{-2}$
\label{t:corel}
\end{table*}

Additionally, introducing $B$ to the global parameters in the subsequent PCA distributes the information
on galaxies essentially into two PC components.
This is  illustrated in Table~\ref{t:pca1}, where the first row gives the 
eigenvalues that measure the variance in the direction of associated PCs. The sum of eigenvalues gives the total 
variance in the data, which in our approach is just the number of PCs, as the original variables were 
standarised. The second row shows the proportion of eigenvalues to the total data variance and determines 
how big a fraction of the total variance is accounted for by the subsequent PCs. The next part of the table 
shows in respective columns the components of eigenvectors associated with individual PCs, which can be understood 
as to what extent each original variable contributed to building a PC.
On examining the values presented in Table~\ref{t:pca1} one can see  that PC1 contains
mostly information from the global parameters, as in the previous analysis, but involves also a contribution 
from some (systematic) part of magnetic field $B$, which is less than in the case of the global parameters. 
In contrast, most of the information about magnetism is independent of the other parameters and constitutes the next
component, PC2. Both PCs account for 75\% and 14\% of the variability in the data, respectively,
which suggests that in this description of galaxies, the global parameters carry much more information than
the magnetic field strength.

In our third approach to PCA, we analysed the intensive parameters (SSFR, SFE,  SFE$_{\mathrm{H2}}$,
$SM_{\mathrm{HI}}$, $SM_{\mathrm{H2}}$, $SLK$). Here, only four (SSFR, SFE, $SM_{\mathrm{H2}}$, $SLK$) out
of six variables significantly contribute to PC1, which accounts for 51\% of the population variability.
The other parameters, SFE$_{\mathrm{H2}}$ and $SM_{\mathrm{HI}}$, dominate the components PC2 and PC3, respectively.

Subsequently, we added information about $B$, which passed almost completely into PC1, where it constituted a factor
comparable to the other intensive parameters (see Table~\ref{t:pca2}). The next two primary components
are dominated again by SFE$_{\mathrm{H2}}$ and $SM_{\mathrm{HI}}$. The first three components combined describe
91\% of the data variance. Contrary to PCA performed on global parameters the magnetic field thus appears
equally important as SSFR, $SM_{\mathrm{H2}}$, and $SLK$, in accounting for the intensive properties of galaxies.

In the final analysis, we took into account all the intensive parameters, including $B$, and the global ones.
From the comparison of eigenvector components, it is clear that the strength of magnetic field
is connected  mainly to the intensive parameters, while the global parameters have only weak relationships 
with $B$. 

This is apparent in the correlation vector diagram (biplot in Fig.~\ref{f:biplot}), 
which shows two-dimensional projections of each data point onto the first two PCs
and the components of eigenvectors (shown as arrows) representing the original variables 
as projected into the PC1-PC2 plane. The elements of the vectors correspond to the correlations 
of each variable with each PC. As the cosines of the angles between the different vectors 
are a measure of correlation between the respective variables, the vectors pointing in 
the same direction represent the perfectly correlated variables, while the perpendicular ones 
indicate a complete lack of correlation. 
In our plot the vector corresponding to $B$ is surrounded solely by the vectors
of intensive parameters, which suggests that they are closely related. The angles between vectors representing
the intensive parameters (including $B$) and the global ones are large, indicating just weak associations.

Galaxies appear to be well grouped in the PC1-PC2 plane (Fig.~\ref{f:biplot}). In particular, the low-mass objects acquire 
the highest value of the component PC1 and are located to the right on the graph. More massive objects exhibiting the strongest
star formation (LIRGs, M\,82) occupy the bottom-left part of the chart and have a small value of the PC1.
The starbursting dwarfs NGC\,1569 and IC\,10 lie between them, while the normal spirals are on the other
side of the plot.

\subsection{Regressions}
\label{s:regressions}

\begin{figure*}[t]
\centering
\includegraphics[clip,width=0.33\textwidth]{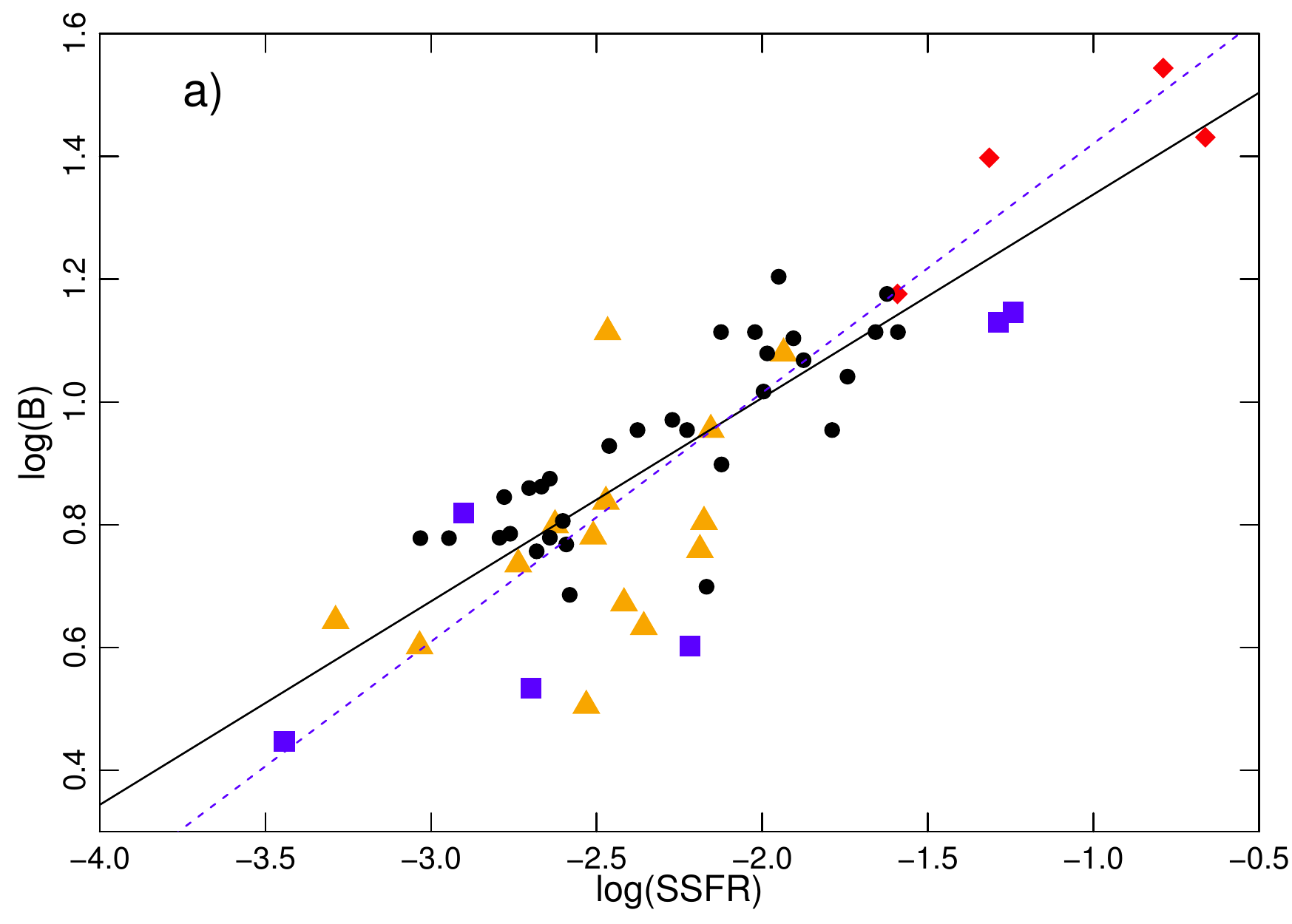}
\includegraphics[clip,width=0.33\textwidth]{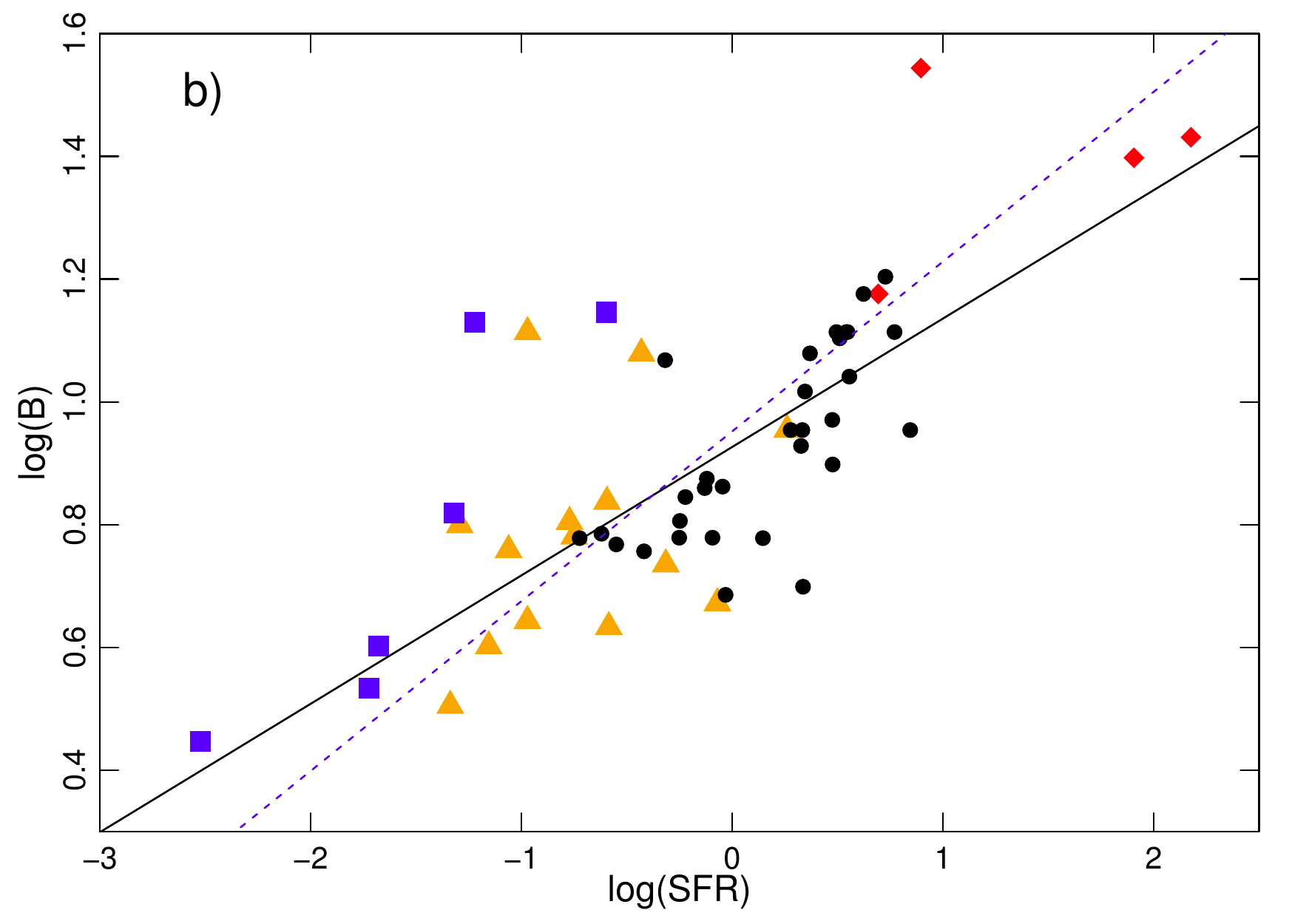}
\includegraphics[clip,width=0.33\textwidth]{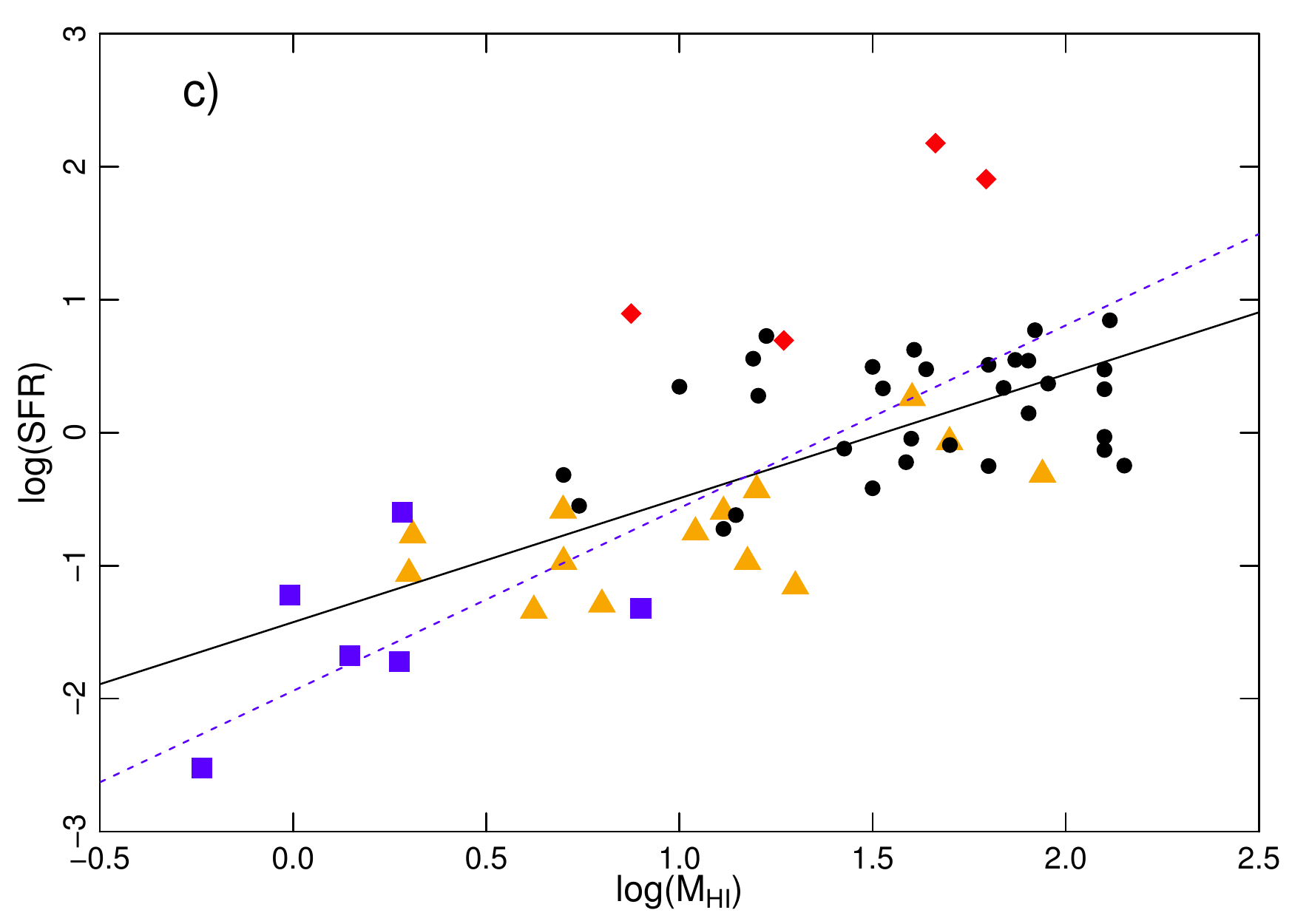}
\includegraphics[clip,width=0.33\textwidth]{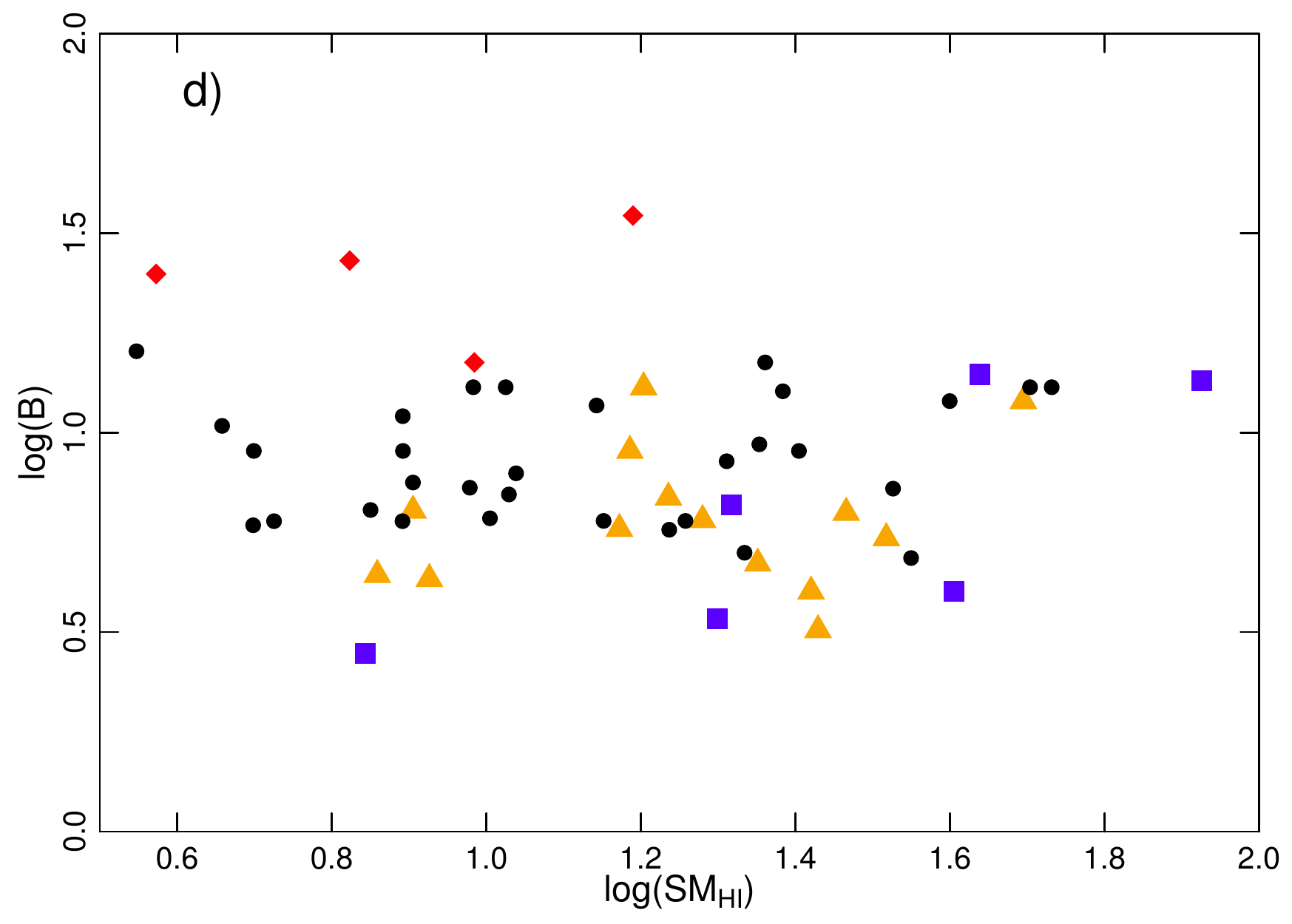}
\includegraphics[clip,width=0.33\textwidth]{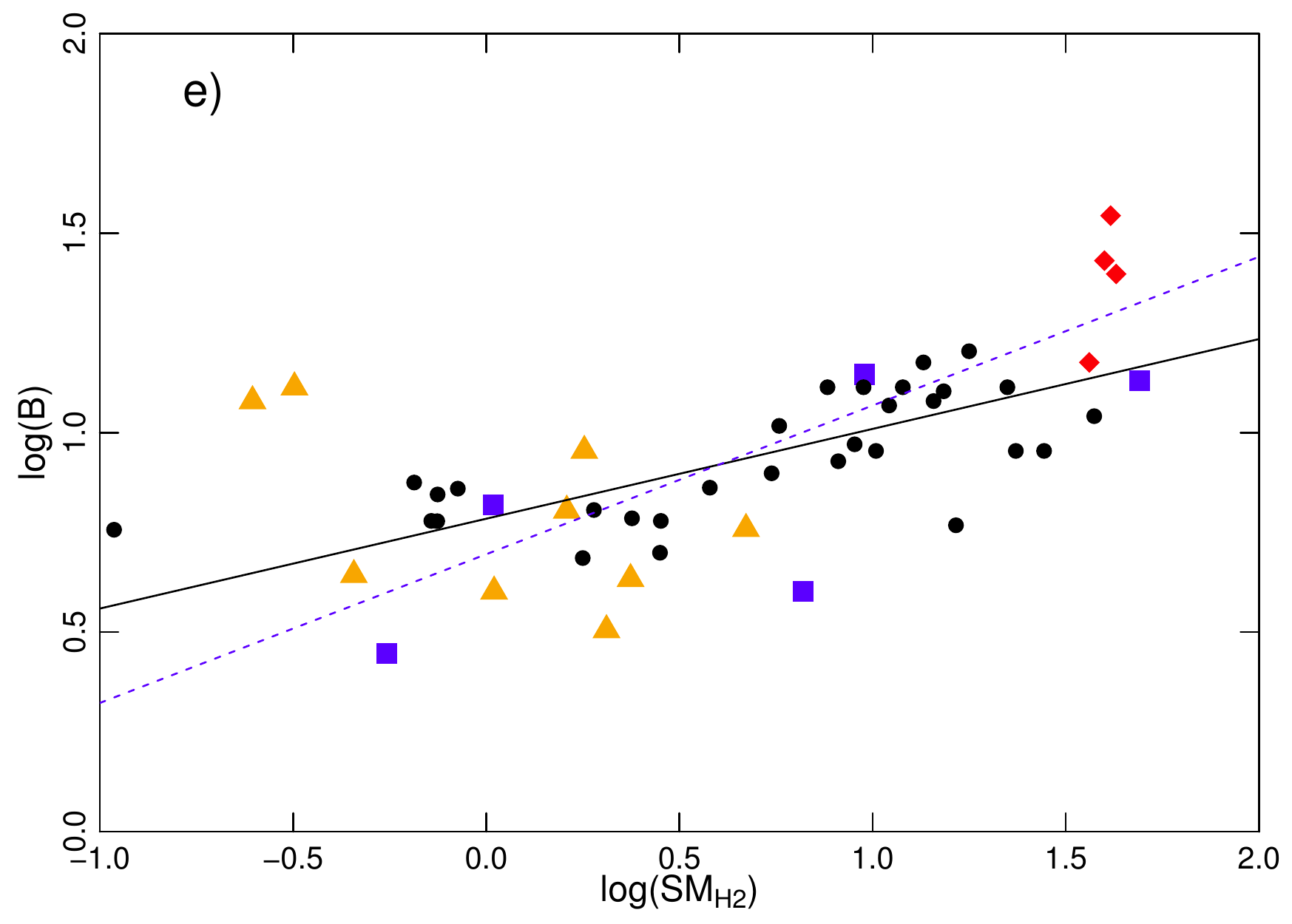}
\includegraphics[clip,width=0.33\textwidth]{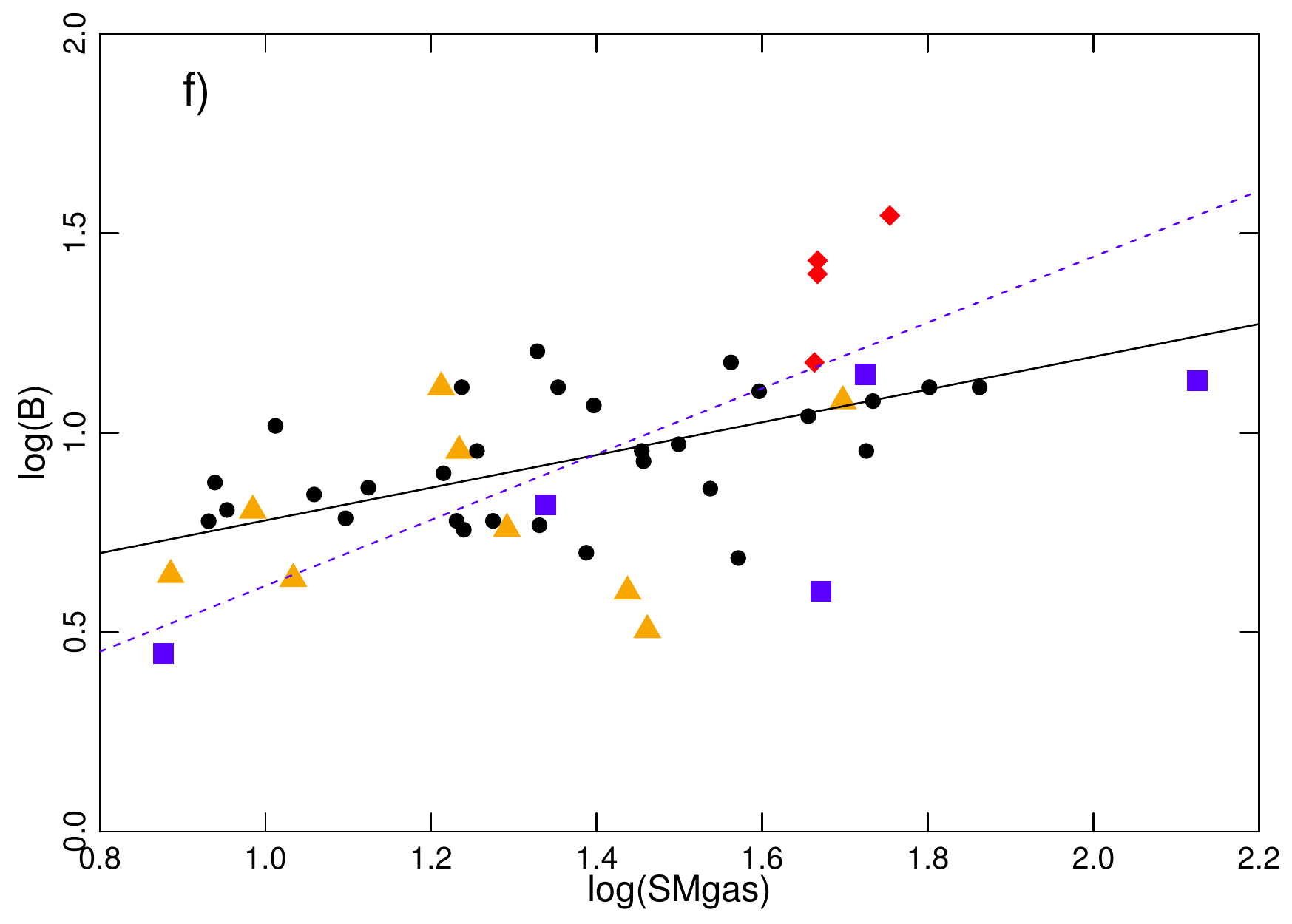}
\includegraphics[clip,width=0.33\textwidth]{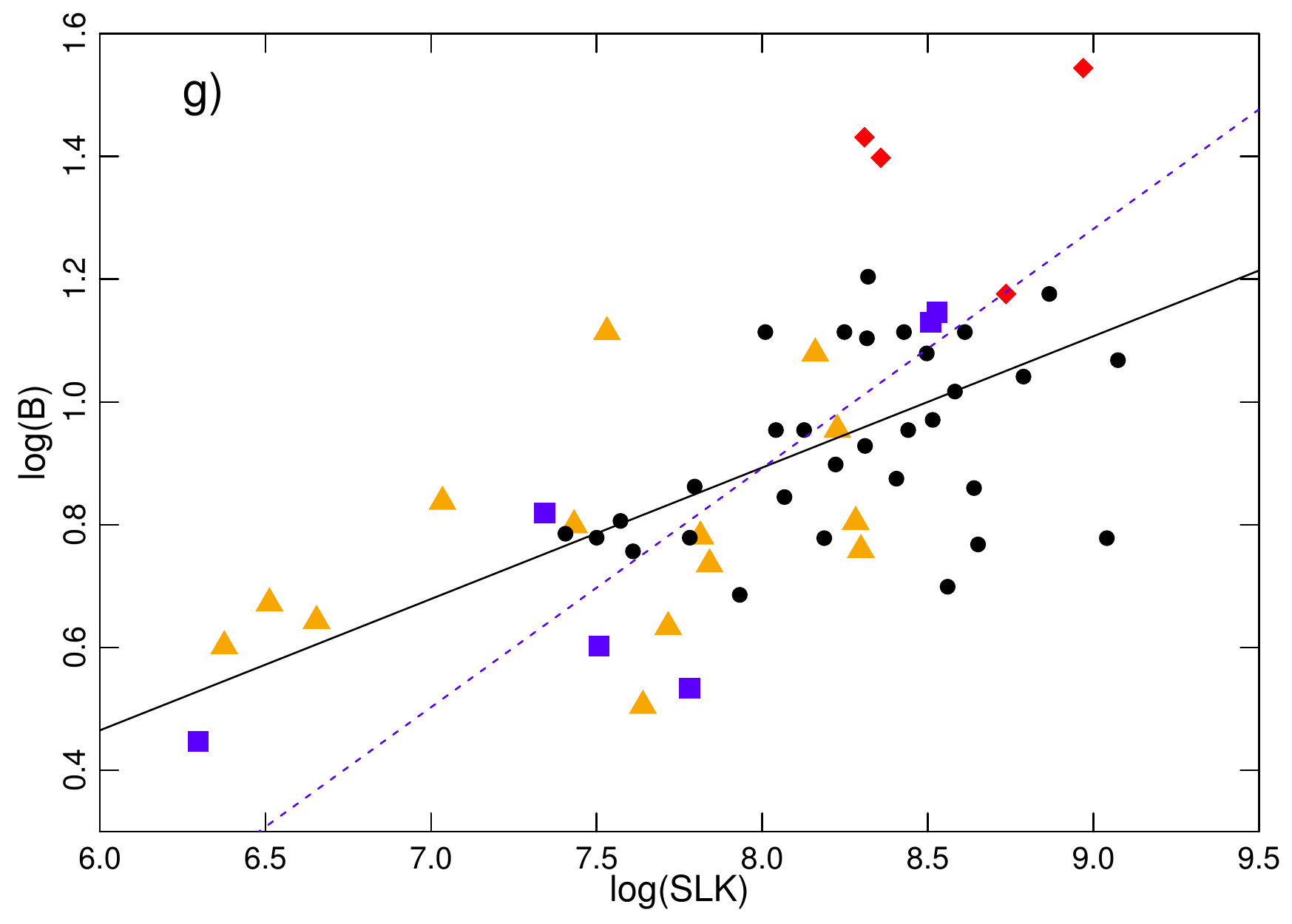}
\includegraphics[clip,width=0.33\textwidth]{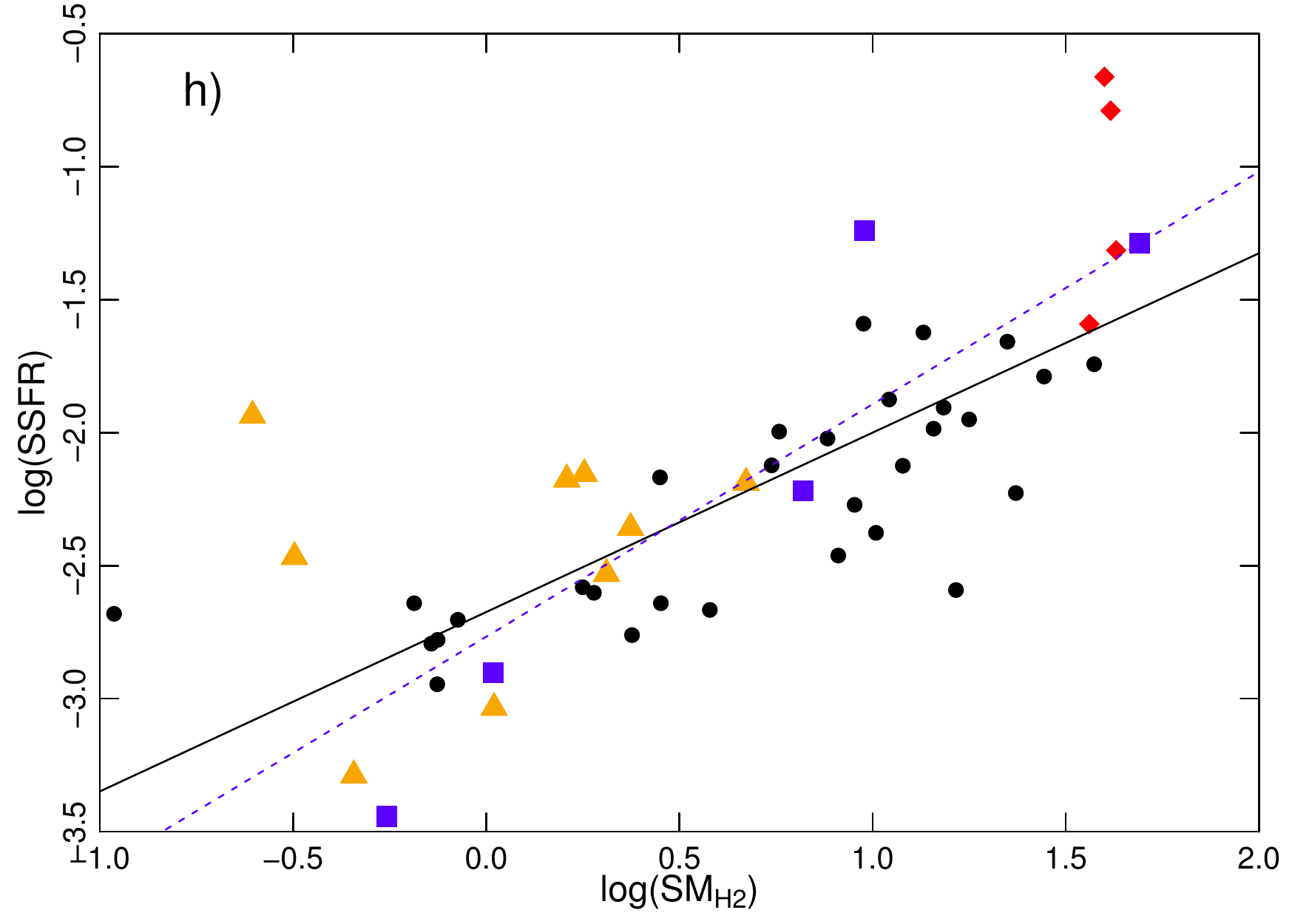}
\includegraphics[clip,width=0.33\textwidth]{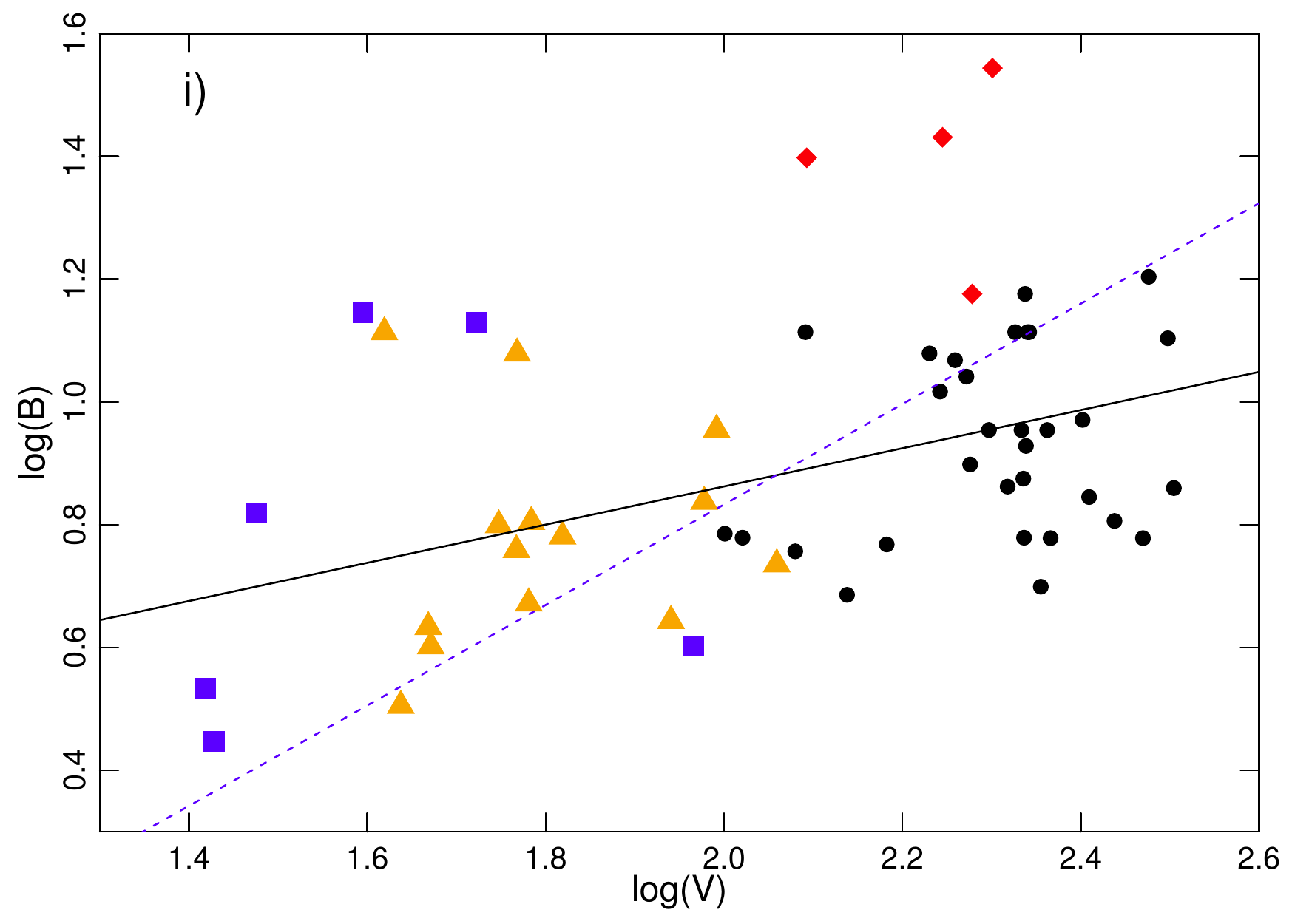}
\caption{Relations between various galaxy parameters for sample galaxies of different categories: 
dwarfs -- rectangles, Magellanic and peculiar low-mass galaxies -- triangles, 
spiral galaxies -- circles, massive starbursts and LIRGs -- diamonds. 
The solid line represents the M-estimation of Y/X regression 
and the dashed line denotes the bisector fit. }
\label{f:corel}
\end{figure*}

The influence of galaxy extensive and intensive properties on  magnetic field can be quantitatively assessed by 
regression methods and expressed in functional form. Following some earlier attempts (e.g. Chy\.zy et al.~\cite{chyzy11}, 
Heesen et al. \cite{heesen14}, Van Eck \cite{vaneck15}, Tabatabaei et al. \cite{taba16}, 
Tabatabaei et al. \cite{taba17}), we approximated the data using power-law functions, which correspond to linear fits
after converting to the logarithmic scale. To remove possible data outliers, we used a robust M-estimation
of two-dimensional (Y/X) regression by the means of iterated re-weighted least squares.
The method was used instead of the ordinary (Y/X) least squares regression, but actually in all cases the results
obtained from both the methods were very similar. We also applied bisector regression, that treats the variables 
in a symmetrical way. For finding the strength of relationship between the parameters, the Spearman's rank 
correlation coefficient was determined.

The most significant correlation found in the relationship of magnetic field with the galaxy parameters
is the relation $B$-SSFR ($\rho=0.78$, Table~\ref{t:corel} and Fig.~\ref{f:corel}a). The fitted index of the 
power law $n=0.33\pm 0.03$  is almost identical with the one obtained for the dwarf irregular galaxies only: $0.30\pm 0.04$ (Chy\.zy
et al. \cite{chyzy11}). The magnetic field is also associated with the global SFR but to a smaller extent ($\rho=0.68$ ,
$n=0.21\pm 0.02$).

We found that the total magnetic field strength $B$ is significantly correlated ($\rho=0.65$) with the surface density of
molecular (H$_2$) gas but not correlated with neutral gas $SM_{\mathrm{HI}}$ ($\rho=-0.03$) (see Figs.~\ref{f:corel}d-e).
As $B$ is closely associated with SSFR, the difference could presumably have arisen from the observed different 
linking of SSFR with the density of neutral gas ($\rho=0.15$) and of molecular gas ($\rho=0.78$) (see Fig.~\ref{f:corel}h).
We checked that $B-SM_{\mathrm{HI}}$ and $B-SM_{\mathrm{H2}}$ relationships for our sample are similar 
to those observed for the sample of Van Eck et al.~(\cite{vaneck15}). 
In our previous work, we found a distinct $B$-$SM_{\mathrm{HI}}$ relation for a group of low-mass (dwarf) galaxies
(Chy\.zy et al. \cite{chyzy11}). This makes for a remarkable difference with our current study. We think that this can
be possibly related to galactic mass (or SFR), since the more massive galaxies we took into consideration, the smaller 
$B-SM_{\mathrm{HI}}$ correlation was observed. When we restricted our sample so as to not include massive starbursts, 
just a weak correlation emerged (Table~\ref{t:corel}). 
The work of Bigiel et al. (\cite{bigiel08}) can further support this view as it shows that $SSFR-SM_{\mathrm{H2}}$ 
relation for {H}{I} dominated dwarf irregular galaxies resemble the coupling found in outer parts of spiral galaxies, 
but galaxies with higher fraction of H$_2$ gas or inner parts of spiral galaxies can show slightly different relationship.
Bigiel et al. received $n=1.0\pm 0.2$ for $SSFR-SM_{\mathrm{H2}}$ relation for galaxies in the regime where 
${SM}_\mathrm{H2}=3-50 \,M_{\sun}\,\mathrm{pc}^{-2}$. When our sample was restricted to this range we obtain 
similar relation with an index of $0.96\pm0.20$.

The relation of $B$ with the total gas density ($SM_\mathrm{gas}=SM_{\mathrm{HI}}+SM_{\mathrm{H2}}$) 
is also statistically significant for our sample showing a power-law index $n=0.41\pm0.10$ and $\rho=0.52$.
We note that within regions in M\,31 $B$ was found even to be best coupled to the volume density of the total gas rather 
than to a specific component (Berhhuijsen et al. \cite{berkhuijsen93}). 
For more H$_2$ dominated galaxies we expect $B-SM_{\mathrm{H2}}$ relation to be the strongest one 
due to a clear, monotonic $SSFR-SM_{\mathrm{H2}}$ relationship shown by Bigiel et al. (\cite{bigiel08}).

For our sample the magnetic field does not show any significant relation with the star formation 
efficiency based on H$_2$ ($\rho=0.07$). We notice strong association of $B$ with the star formation efficiency 
based on neutral gas (SFE) with $\rho=0.75$, but this did not provide us with any new information. We explain this 
association as a result of the mentioned strong $B$-SSFR correlation and the lack of significant relationships between 
SSFR and $SM_{\mathrm{HI}}$ (Table~\ref{t:corel}). 

The comparison of the strength of correlation of $B$ with global $M_{\mathrm{HI}}$, $M_{\mathrm{H2}}$, and $M$, shows 
that $B$ is not closely connected with the total mass $M$, which is the largest source of gravitational force. 
Actually, the strongest relation occurs with the molecular mass - that part of the galactic mass which 
is most related with production of stars. Therefore, we interpret the dependence of $B$ on $M$ (as well as on $V$) as 
an indirect one, resulting from the $B$-SFR coupling and the observed connection of global SFR with the available 
total molecular mass in galaxies. The association we find between $B$ and $LK$ ($\rho=0.49$), a rough 
estimator of stellar mass in galaxies and stellar activity, may support
this line of reasoning.

We checked whether the galaxy inclination is related to any other parameters and whether it 
could have affected our results. The calculated correlation coefficient between the inclination and 
the other parameters turned out to be statistically non-significant. We then applied a simple correction 
for the inclination in calculating the surface density of the SFR: instead of the galaxy observed 
surface area, we scaled the SFR by the area of a circle with the radius equal to the galaxy 
major axis. We repeated the regression analysis for the magnetic field and thus obtained 
the surface density of the star formation rate $\mathrm{SSFR_{cor}}$. The fitted index of the power law 
$n=0.31\pm 0.03$ is very similar to the original one (Table~\ref{t:corel}), which proves once again that 
inclination does not change the calculated relationships by more than  statistical uncertainties. 

\section{Discussion and conclusions}
\label{s:discussion}

The PCA allowed us to compare the significance of relations of $B$ with various galaxy parameters, demonstrating 
that the global galaxy parameters are all mutually correlated and can be represented by a single principal 
component. Thus our sample reproduces the result of Disney et al. (\cite{disney08}), who had used
almost 200 galaxies (Sect.~\ref{s:intro}). According to our analysis, the values of magnetic field are
not too closely related to the global parameters, hence the latter cannot be a major drivers of magnetic fields.
Nevertheless, the PCA and regression analysis do reveal weak correlations of $B$ with the global parameters,
for example, the global SFR (Sect.~\ref{s:regressions}).

\begin{figure}
\centering
\includegraphics[clip,width=0.50\textwidth]{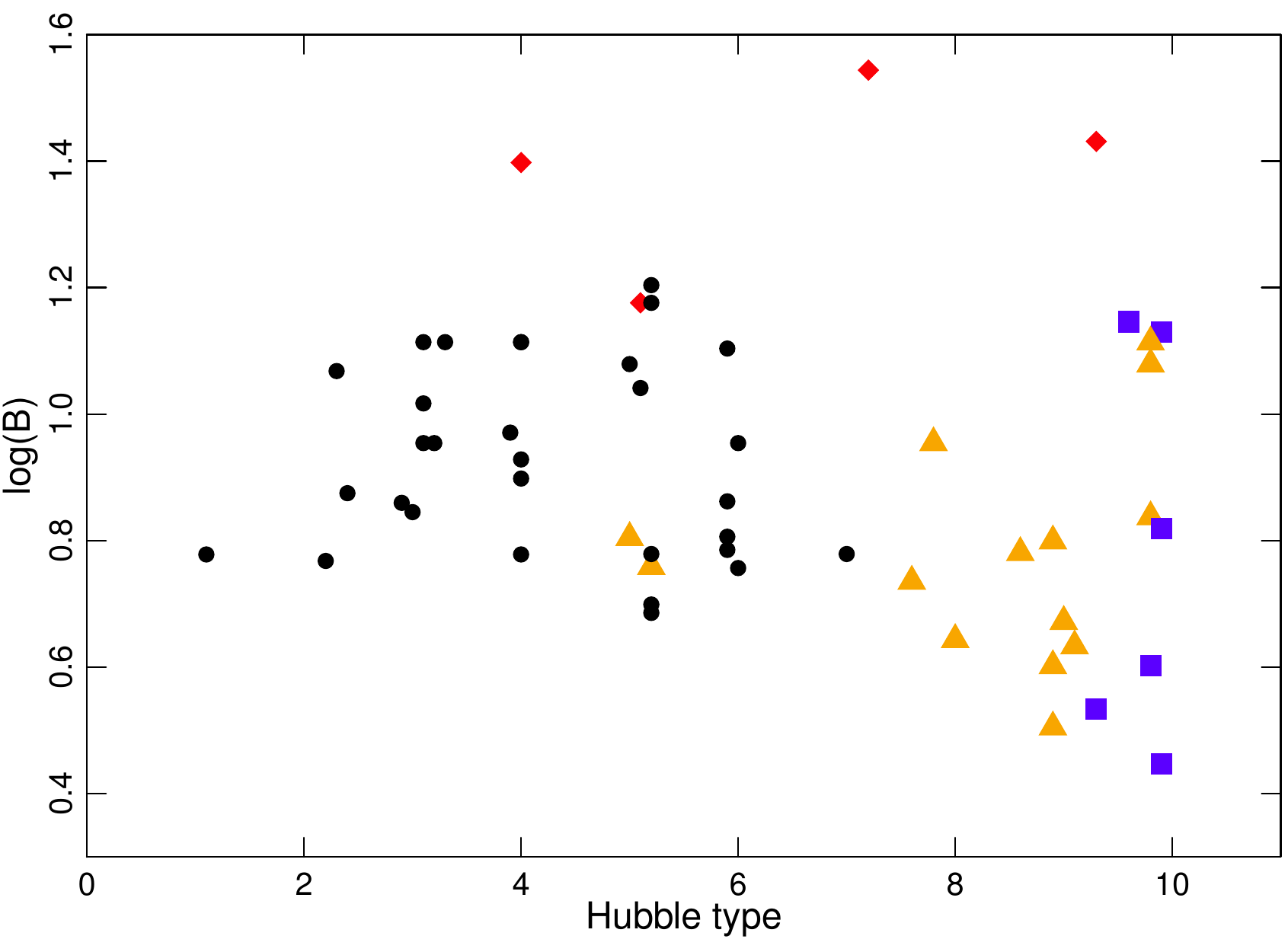}
\caption{Magnetic field strength $B$ along the Hubble sequence.  
Symbolic markers are the same as in Fig.~\ref{f:corel}.
}
\label{f:hubble}
\end{figure}

In order to probe these connections and spotting in Fig.\ref{f:corel}a that the locations of galaxies depend on
their category, we constructed a graph of $B$ along the Hubble sequence (Fig.\ref{f:hubble}). There is a large 
diversity of observed strengths of magnetic field for almost each Hubble type. The maximum values of $B$ are 
not restricted by the morphological type and even dwarf galaxies (those which are in the starburst phase) are 
able to produce strong total magnetic fields.
However, it can be noticed that the lower envelope of field strength varies with the type in a systematic way. Weaker 
fields appear exclusively in later Hubble types ($T>8$) and the mean strength as low as about $5\,\mu$G is not observed
in the normal spiral galaxies. We suspect these differences are due to density waves, which in the typical spiral galaxies
always force some minimal level of star forming activity and in turn, subsequent production of magnetic fields by 
the small-scale dynamo.

We also notice relatively weak fields for early types of galaxies (Fig.~\ref{f:hubble}), although this part of the diagram
requires more data to verify this observation. A systematic decrease of $B$ towards the early-type galaxies is expected:
in the Sa ($T=1$) galaxies, massive stars form usually in small clusters, while in the Sc-d ($5\le T \le 7$)
objects \ion{H}{II} associations containing hundreds or thousands of OB stars are found (Kennicutt \cite{kennicutt98b}).
As the stellar activity modifies the structure and dynamics of ISM, we can suppose that magnetic field topologies
and strengths are accordingly changed and weaker fields occur in more quiet ISM.

We find that the closest relationship of $B$ is with SSFR ($\rho=0.78$), which is described by a power-law with
an index $n=0.33\pm 0.03$. As this relation is in excellent correspondence to the one determined for 
low-mass galaxies alone ($0.30\pm 0.04$, Chy\.zy et al. \cite{chyzy11}) it shows that the processes of generating 
magnetic field in the dwarf and Magellanic-type galaxies are similar to those in the massive spirals. 
In the present analysis the statistical sample (of 55 objects) is several times larger than the previous 
one and not only supports but also even strengthens the results obtained from the Local Group dwarfs.
This trend is observed over three orders of magnitude in SSFR for galaxies, while the global SFR spreads over 
more than four orders of magnitude. Also the three starburst galaxies with highest SSFR (Fig.~\ref{f:corel}a) 
fit the trend. Hence, we can reasonably suspect that the distant galaxies with extremely high SFR (like Ultra LIRGs) 
would also follow this relationship. Deep radio surveys, for example with LOFAR (Hardcastle et al. \cite{hardcastle16}), 
can potentially provide appropriate observational evidence.

Our sample is large enough to statistically compare for the first time the production levels of magnetic fields 
in the spirals, dwarf and irregular galaxies having similar SSFR. The relevant data can be seen in the 
categorial plot of $B$ against SSFR in Fig.~\ref{f:corel}a. It appears that the spiral galaxies have slightly 
stronger fields than dwarfs (in agreement with Fig.~\ref{f:hubble}). Different galaxy mixes can thus lead to 
different power-law indices in $B$-SSFR relation, which may explain the slightly different results 
reported in previous published works (see e.g. Van Eck et al. \cite{vaneck15} and Sect.~\ref{s:intro}).

In our sample, the total magnetic field is correlated with the density of cold molecular (H$_2$) gas but not with 
warm neutral (\ion{H}{I}) medium (Figs.~\ref{f:corel}d-e). This is supported by similar results obtained by Bigiel et 
al. (\cite{bigiel08}) and Van Eck et al.~(\cite{vaneck15}) for different galaxy samples. According to our work, 
a best-fit Schmidt law (SSFR-${SM}_\mathrm{gas}$) shows an exponent $n=1.39\pm 0.20$ (Table~\ref{t:corel}) whereas 
Kennicutt (\cite{kennicutt98a}) found $n=1.40\pm 0.15$ for actively star-forming galaxies. 

Considering the above we propose two possibilities to simply interpret the observed $B$-SSFR relation.
According to the first idea this relation partly results from a tight correlation between 
radio luminosity $LR$ and the infrared luminosity which is closely connected to the global SFR. 
Modelling of this relation which can be described by a power-law with an index $\beta$ 
($LR \propto SFR^\beta$) usually assumes proportionality of radio luminosity to the CRs production rate, 
which itself is proportional to the supernova rate and hence to the SFR. Different galaxy properties and 
environment may further involve other processes as CRs and dust-heating UV-photons escape or synchrotron 
emission from secondary CR electrons produced by interaction of CRs protons with dense molecular clouds. 
Assuming further energy equipartition between magnetic fields and CRs yields a formula for radio intensity 
$I \propto B^{3+\alpha}$ where $\alpha$ is the radio spectral index, which allow us to re-write the radio-infrared 
relation to the form: $B \propto \mathrm{SSFR}^{\beta/(3+\alpha)}$. The observed relation $B\propto \mathrm{SSFR}^{0.33}$ 
and typical value of $\alpha=0.9$ results in the radio-infrared relation with $\beta=1.29$. 
This value is in a good agreement with observations (see e.g. Heesen et al. \cite{heesen14}, Beck \cite{beck16}). 

The second interpretation of the $B$-SSFR relation we base on the SSFR-$SM_{\mathrm{gas}}$ coupling 
(the Schmidt law with the observed exponent $n=1.39\pm0.20$) which leads to $B \propto SM_{\mathrm{gas}}^{0.46}$. 
Then we assume turbulent magnetic field amplification, for example by a small-scale dynamo, which results in scaling 
of the magnetic energy with the turbulent energy of the gas: $B^2 \propto SM_\mathrm{gas} v^2$ where 
$v\approx10$\,km\,s$^{-1}$ is the turbulent gas velocity. This results in scaling $B\propto SM_{\mathrm{gas}}^{0.5}$ 
which well corresponds with the derived exponent 0.46 and the observed exponent $0.41\pm 0.10$ (Table~\ref{t:corel}). 
More detailed description of physical processes involved in amplification of magnetic fields by a small-scale 
dynamo by Schleicher \& Beck (\cite{schleicher13}) leads to the relationship ($B\propto \mathrm{SSFR}^{1/3}$) 
which is very similar to the observed one. 

The results from the stack experiment involving the radio-faint dwarf galaxies (the `common' sample, 
Sect.~\ref{s:lowmasssample}) from Roychowdhury \& Chengalur (\cite{roy12}), can also be compared with our 
$B$-SSFR relation. The value $B=1.4\,\mu$G and $\mathrm{SSFR}=9.8\times 10^{-4}\,\mathrm{M_{\sun}\, yr^{-1}\, kpc^2}$ 
locates these objects significantly ($\approx 1\,\mu$G) below the trend (Fig.~\ref{f:corel}a). This difference 
is not likely due to errors. The value of $B$ is lower than the magnetic field equivalent, due to inverse Compton losses 
of relativistic electrons in the cosmic microwave background. Hence, the strength estimated from the presumably 
reduced synchrotron emission can be undervalued. Additionally, at such low SSFR the turbulence injection timescale 
(or timescale of massive star formation) can become longer than the dissipation timescale of CR electrons and brake 
the equipartition between magnetic fields and CRs resulting in decrease in synchrotron emission and $B$ 
(see Schleicher \& Beck \cite{schleicher16}).

In the case of faint, radio-undetected dwarf galaxies of the Local Group, instead of using results from
the stack experiments (Sect.~\ref{s:lowmasssample}), we take for the purpose of analysis the upper limit of $B=4\,\mu$G from
Chy\.zy et al. (\cite{chyzy11}) and determine $\mathrm{SSFR}=7.3\times 10^{-5}\,\mathrm{M_{\sun}\, yr^{-1}\, kpc^2}$
from the data presented in that work. The obtained position for these dwarfs is deflected slightly above the global $B$-SSFR trend.
Therefore, these objects and those from the `common' sample were not included in other statistical analyses.

Differential rotation and large-scale dynamo are indispensable to account for the ordered part of magnetic 
field in galaxies. In the work of Tabatabaei et al. (\cite{taba16}), only the ordered part of magnetic field was investigated
for a sample of 26 galaxies and found to be correlated with the dynamic mass and the rotational velocities of galaxies.
In our sample, only the total field was analysed, but it also showed the relationships with $V$ and $M$ of roughly similar
strength ($\rho=0.30-035$). As the ordered field contributes just little to the total field, the argument of Tabatabaei 
et al. (\cite{taba16}) that the massive, faster-rotating galaxies compress and share turbulent magnetic field leading 
to stronger ordered fields is not valid for our $B-M$ and $B-V$ relations (see Fig.~\ref{f:corel}i). 
As shown in Sect.~\ref{s:regressions}, the total magnetic field $B$ in our objects is strongly associated with 
the star formation rate ($\rho=0.68$), and even more strongly with the SSFR ($\rho=0.78$). 
Such relationships can be explained by the turbulent energy injected to the ISM through supernova explosions 
and amplification of magnetic fields by a small-scale dynamo (Schleicher \& Beck \cite{schleicher16}). Hence, we suspect 
that $B$ is directly related to the SSFR or $SM_{\mathrm{H2}}$, while, since the amount of molecular gas available for star formation 
is related to the total mass of galaxies (Sect.~\ref{s:regressions}), the relation of $B$ with galactic 
mass or rotation is only an indirect one.

In our sample, the $B$-SSFR relation is also fulfilled by dwarf galaxies and massive starbursts, which usually 
manifest slow or disordered rotation. We have shown that even dwarf galaxies with slow rotation and low mass 
(as e.g. IC\,10) can develop strong magnetic fields in the starburst phase. Therefore, for our sample of galaxies, 
it is  the small-scale dynamo mechanism rather than the large-scale one that decisively determines the magnetic field strength. 

We note that some relations beween $B$ and intensive variables presented throughout this work could be stronger if they were
determined only over the regions of high star-forming activity. In our approach, we applied the average values, based on 
the full extent of galaxies. Further investigation of these different approaches involving a larger sample of galaxies, 
from the upcoming large area radio continuum survey with the LOFAR (Shimwell et al.~\cite{shimwell16}) and the APERTIF 
(Verheijen et al.~\cite{verheijen09}) radio telescopes, are highly desirable.

\begin{acknowledgements}
This research was supported by Polish National Science Centre through grant 2012/07/B/ST9/04404.
We thank Dr.~Rainer Beck and the anonymous referee for the detailed and constructive comments. 
We acknowledge the use of the HyperLeda (http://leda.univ-lyon1.fr) and NED 
(http://nedwww.ipac.caltech.edu) databases.
\end{acknowledgements}

\begin{appendix}
\section{Information on galaxies}

\begin{table*}[t]
\caption{Breakdown of basic properties of the galaxy sample by category.
}
\centering
\begin{tabular}{lccccccccc}
\hline\hline
Galaxy & Hubble & $B$     & SFR & $M_{\mathrm{HI}}$ & $M_{\mathrm{H2}}$ & $LK$ & $R$ & $V$ & References \\ 
       & type $T$   & $\mu \mathrm{G}$ & $\mathrm{M_{\sun}}\,\mathrm{yr}^{-1}$ & $10^{8}\,\mathrm{M_{\sun}}$ & $10^{8}\,\mathrm{M_{\sun}}$ & $10^{8}\,\mathrm{erg\,s}^{-1}$ & kpc & $\mathrm{km\,s}^{-1}$ & for Columns\\
 1 & 2 & 3     & 4  & 5 & 6 & 7 & 8 & 9 & 3-7, 9 \\       
\hline
Dwarf and Magellanic-type       &               &               &       &       &       &               &       &       &       \\\hline
NGC292(SMC)     &8.9    &       3.2     &       0.05    &       4.2     &       0.3     &               6.8     &               2.9     &       43&     1,      1,      1,      40,     50,     16 \\
NGC1569 &       9.6     &       14.0    &       0.25    &       0.6     &       0.4     &               14      &       1.8     &       39&     1,      12,     23,     58,     24, 24\\
NGC2976 &       5.2     &       5.7&            0.09    &       2.0     &       0.6     &               26      &       3.1     &       58& 1, 13, 13, 13, 24       , 24    \\
NGC3239 &       9.8     &       6.9&            0.25    &       13      &N/A    &               8.2     &               6.0     &       95&     1, 1, 1,    - , 24, 24\\
NGC4027 &       7.8     &       9.0     &       1.82    &       40      &4.7&           439     &       10              &98& 1, 1, 1, 41, 24, 24     \\
NGC4214 &       9.8     &       13.0    &       0.11    &       5.0 &   0.1     &               10      &       3.6     &       42&     54, 13, 13, 13, 24, 13\\
NGC4236 &       8.0     &       4.4&            0.11    &       15      & 0.9&            9.4     &               14              &87& 1, 1, 1, 55, 24, 51  \\
NGC4449 &       9.8     &       12.0    &       0.37    &       16      &0.1&           46      &       3.8     &       59&     1, 13, 13, 13, 49, 24\\
NGC4605 &       5.0     &       6.4     &       0.17    &       2.0     &       0.4     &               48      &       4.6     &       61&     1, 1, 1, 56, 24, 24\\
NGC4618 &       8.6     &       6.0     &       0.18    &       11      &N/A    &               37      &       4.8     &       66&     1, 1, 1, -, 24, 24\\
NGC4656 &       9.0     &       4.7     &       0.85    &       50      &N/A    &               7.2     &               18              &60 &       1, 1, 1, -, 24, 24\\
NGC5204 &       8.9     &       6.3     &       0.05    &       6.3     &       N/A     &               5.8     &               3.4     &       55&     1, 1, 1, -, 24, 24\\
NGC6822 &       9.8     &       4.0     &       0.02    &       1.4     &       0.2     &               1.1     &               1.1     &       92 &       1, 1, 1, 42, 49, 26\\
UGC11861        &       7.6     &       5.4     &       0.48    &       87      &N/A    &               183     &       10              &114&   1, 1, 1, -, 24, 24\\
UGC5456 &       9.3     &       3.4     &       0.02    &       1.9 &   N/A     &               5.7&            2.5     &       26&     10, 14, 24, -, 24, 24\\
HoII    &               9.9     &       6.6     &       0.05    &       7.9     &       0.4     &               8.4&            3.9     &       29&     4, 13, 13, 13, 24, 13\\
IC10    &               9.9     &       13.5    &       0.06    &       0.9     &       0.6     &               3.8&            0.7     &       52&     5, 15, 1, 42       , 49    , 16\\
IC1613  &       9.9     &       2.8     &       $<$0.01 &0.6    &       0.1     &               0.2     &               1.7     &       26&     1, 1, 1, 42, 50, 16\\
IC2574  &       8.9     &       4.0     &       0.07    &       19      &0.8    &       1.8     &               7.7     &       46&     1, 13, 13, 13, 50, 24\\
LMC     &               9.1     &       4.3&    0.26    &       5.0     &       1.4     &               31      &       4.7     &       46&     1, 1, 1, 43, 50,    16\\\hline
Spiral  &               &               &       &       &       &               &       &       &       \\\hline
NGC0224(M31)    &3.0    &       7.0     &       0.60    &       39      &2.7    &               421     &       19         &256&   11, 16, 25, 57, 49, 16\\
NGC598(M33)     &5.9    &       6.1     &       0.24    &       14      &3.3    &       35         &       8.7     &       100&    11, 15, 26,     26,     49,     16\\
NGC628  &       5.2     &       6.0     &       0.81    &       50      &10     &       213     &       11         &217&   4,      13,     13,     13,     49,     13\\
NGC891  &       3.1     &       13.0    &       3.48    &       80      &35     &       647     &       16         &212&   11,     17,     27,     23,     49,     16\\
NGC925  &       7.0     &       6.0 &   0.56    &       63      &2.5    &       110     &       14         &104&   4,      13,     13,     13,     24,     13\\
NGC1097 &       3.3     &       13.0    &       5.90    &       83      &94     &       1390         &       19      &219&   11,     12,     28,     44,     49,     16\\
NGC1365 &       3.2     &       9.0     &       7.00    &       130     &170    &       2229         &       31      &198&   11,     16,     30,     30,     49,     16\\
NGC1566 &       4.0     &       13.0    &       3.53    &       74      &13     &       140     &       7.4     &       123&    11,     12,     31,     52,     24,     31\\
NGC2403 &       6.0     &       5.7     &       0.38    &       32      &0.2    &       74 &       10       & 120& 4,      13,     13,     13,     49,     13\\
NGC2841 &       2.9     &       7.2     &       0.74    &       126     &3.2    &       1633         &       16       & 319& 4,      13,     13,     13,     49,     13\\
NGC2903 &       4.0     &       7.9     &       3.00    &       44      &22     &       662         &       16      &188&   4,      8,      8,      8,      49,     24\\
NGC3031(M81)    &2.4    &       7.5     &       0.76    &       27      &2.2    &       844         &       14      &216&   11,     18,     32,     23,     49,     24\\
NGC3184 &       5.9     &       7.2     &       0.90    &       40      &16     &       261         &       11      &208&   4,      13,     13,     13,     24,     13\\
NGC3198 &       5.2     &       4.9     &       0.93    &       126     &6.3    &       303     &       17      &137&   4,      13,     13,     13,     24,     13\\
NGC3627 &       3.1     &       10.4    &       2.22    &       10      &13     &       838     &       12      &174&   4,      13,     13,     13,     49,     13\\
NGC3628 &       3.1     &       9.0 &   2.15    &       34      &37     &       365     &       14      &215&   9,      12,     36,     23,     49,     16\\
NGC3992 (M109)&4.0&     6.0     &       1.40    &       80      &N/A    &               2319         &       27      &295&   11, 21, 24,     -,      49,     24\\
NGC4254 (M99)&5.2&      16.0    &       5.34    &       17      &85     &       993     &       13      &299&   11,     12,     23,     23,     24,     24\\
NGC4414 &       5.2     &       15.0    &       4.20    &       41      &24     &       1297     &       10      &217&   11,     19,     24,     -,      24,     19\\
NGC4594 (M104)&1.1&     6.0     &       0.19    &       13      &       0.1     &       1831         &  11   &       232 &   11, 12, 37,     55,     49,     29\\
NGC4736 (M94)&2.3&      11.7    &0.48   &5.0    &       3.9     &       428     &   3.3   &       181 &   4, 13,  13,     13,     49,     16\\
NGC4826(M64)    &2.2    &       5.9     &       0.28    &       5.5&    18      &       494     &       8.1     &       152 &       8, 12,  8,      8,      49,     24\\
NGC5055 &       4.0     &       8.5     &       2.12    &       126     &50     &       1257         &       18      &218 &  4,      13,     13,     13,     49,     13\\
NGC5194(M51)    &4.0    &       13.0    &       3.13    &       32      &25     &       881     &       13      &219  &      4, 13,  13,     13,     49,     13\\
NGC5236(M83)    &5.0    &       12.0    &       2.34    &       90      &32     &       710     &       8.9&    170 &       7, 12, 13,      23,     49,     22\\
NGC5457(M101)&5.9&      6.4     &       0.57    &       142     &38     &       747     &       25      &274 & 3,    3,      38,     45,     49,     38      \\
NGC5775 &       5.1     &       11.0    &       3.60    &       16      &75     &       1224    &       16      &187 &       11,     20,     23,     23,     24,     24\\
NGC5907 &       5.2     &       5.0     &       2.17    &       69      &9.0    &       1160    &       30      &226 &       11,     17,     35,     35,     49,     16\\
NGC6946 &       5.9     &       12.7    &       3.24    &       63      &40     &       540     &       9.9&    314 &       4, 13,  13,     13,     49,     16\\
NGC7331 &       3.9     &       9.4&            2.99    &       126     &50     &       1825    &       22      &252 &       4, 13,  13,     13,     49,     13\\
IC342   &       6.0     &       9.0     &       1.89    &       16      &75     &       352     &       10      &230 &       11,     12,     39,     23,     49,     16
\\\hline
Massive starburst/LIRG  &               &               &       &       &       &               &       &       &       \\\hline
NGC253  &       5.1     &       15.0    &       4.94    &       19      &70     &       1051    &       15      &       189 &       11,     16,     28,     23,     49,     16\\
NGC3034 (M82)&7.2&      35.0    &       7.87    &       7.5     &       20      &       451     &       6.3     &               200 &       2,      12,     32,     47,     49,     53\\
NGC3256 &       4.0     &       25.0    &       80.7    &62     &710    &       3793    &       30      &       123 &       6,      6,      33,     48,     24,     33
\\
ARP220  &       9.3     &       27.0    &       150     &46     &275    &       1407    &       17&         175 &        6, 6, 34, 46, 24, 24\\\hline

\hline
\end{tabular}
\label{t:basic}
\begin{flushleft} Notes: Data for Col. 2, and 8 are from HyperLeda and NED. 
References: (1) Jurusik et al.~\cite{jurusik14}, (2)     Adebahr et al.~\cite{adebahr13}
(3)      Berkhuijsen ~\cite{berkhuijsen16}, (4)         Braun et al. ~\cite{braun07}, (5)     Chy\.zy et al.~\cite{chyzy16}, (6)      Drzazga et al. ~\cite{drzazga11}, (7) Neininger et al. ~\cite{neininger93}, (8)   Heesen et al. \cite{heesen14}, 
(9)      Nikiel - Wroczy\'nski et al. ~\cite{nikiel13}, (10) this paper, (11) Van Eck et al. ~\cite{vaneck15}, (12) Calzetti et al. ~\cite{calzetti10}, (13)  Leroy et al. ~\cite{leroy08}, (14) Roychowdhury et al. ~\cite{roy12}, (15) Woo et al ~\cite{wo08}, (16)       Tabatabaei et al. ~\cite{taba16}, (17) Misiriotis et al. ~\cite{misiriotis01}, (18) Karachentsev et al. ~\cite{kara07}, (19) de Blok et al. ~\cite{deblock14}, (20)  Irvin  ~\cite{irvin94}, (21)       Martinet \& Friedli ~\cite{martinet97}, (22)     Heald et al. ~\cite{heald16}, (23) Liu et al. ~\cite{liu15}, (24) LEDA, (25) Cram et al. ~\cite{cram80}, (26) Gratier et al. ~\cite{gratier10}, (27) Sancisi \& Allen ~\cite{sancisi79}, (28)    Koribalski et al. ~\cite{koribalski04}, (29) van der Marel et al. ~\cite{marel94},
(30)    Lindblad ~\cite{lindblad99}, (31) Pence et al. ~\cite{pence90}, (32) Chynoweth et al. ~\cite{chynoweth08},
(33)     English et al. ~\cite{english03}, (34) Baan et al. ~\cite{baan87}, (35)    Dumke et al. ~\cite{dumke97},
(36) Huchtmeier et al. ~\cite{huchtmeier85}, (37)       Bajaja et al. ~\cite{bajaja84}, (38)    Walter et al. ~\cite{walter08}, (39)    Rots ~\cite{rots79}, (40)       Leroy et al. ~\cite{leroy07}, (41)    Casasola et al. ~\cite{casasola04}, (42)        Mateo  ~\cite{mateo98}, (43) Cohen et al. ~\cite{cohen88}, (44) Crosthwaite ~\cite{crosthwaite01}, (45)    Kenney et al. ~\cite{kenney91}, (46)    Papadopoulos et al. ~\cite{papa12}, (47)    Young \& Scoville ~\cite{young84}, (48)         Sargent et al. ~\cite{sargent89}, (49) Jarrett et al. ~\cite{jarrett03}, (50)     NED, (51) Chy\.zy et al. ~\cite{chyzy11}, (52) Combes et al. ~\cite{combes14}, (53) Sofue et al. ~\cite{sofue99}, (54)     Kepley et al. ~\cite{kepley11}, (55) Wilson et al.  ~\cite{wilson12}, (56) Throson \& Bally  ~\cite{throson87}, (57) Dame et al. ~\cite{dame93}, (58) Israel  ~\cite{israel88}

\end{flushleft}
\end{table*}

\end{appendix}

\end{document}